\documentclass[conference]{IEEEtran}[en]
\usepackage{cite}
\usepackage{graphicx}
\usepackage{array}
\usepackage{listings}
\usepackage{url}
\usepackage{caption}
\usepackage{listings}
\usepackage{placeins}
\usepackage{multirow}
\usepackage{lipsum}
\usepackage{multicol}
\usepackage{graphicx}
\usepackage{hhline}
\usepackage{color}
\usepackage{makecell}
\usepackage{xcolor}
\usepackage{gensymb}
\usepackage{subfloat}
\usepackage{tabularx}
\usepackage{comment}
\usepackage{graphicx}
\usepackage{listings}
\usepackage{xcolor}
\usepackage{comment}
\usepackage{diagbox}
\usepackage{makecell}
\usepackage{tikz}
\usepackage{balance}
\usepackage{longtable}
\usepackage{subfig}
\usepackage{amsmath}
\usepackage{mathtools, nccmath}
\usepackage{cuted}
\usepackage{float}

\newcommand{\ie}{\textit{i}.\textit{e}.,\ }
\newcommand{\eg}{\textit{e}.\textit{g}.,\ }
\newcommand{\cf}{\textit{c}\textit{f}.\ }
\newcommand{\1}{\textit{(a)}}
\newcommand{\2}{\textit{(b)}}
\newcommand{\3}{\textit{(c)}}

\newcommand{\solution}{\textit{RCVaR}}

\begin{document}

\title{RCVaR: an Economic Approach to Estimate Cyberattacks Costs using Data from \\ Industry Reports} 

\author{\IEEEauthorblockN{
        Muriel Figueredo Franco, Fabian K\"unzler, Jan von der Assen, Chao Feng, Burkhard Stiller}

\IEEEauthorblockA{
    Communication Systems Group CSG, 	Department of Informatics IfI, University of Zurich UZH\\
	Binzm{\"u}hlestrasse 14, CH--8050 Z{\"u}rich, Switzerland\\
	\\
	E-mail: \{franco, vonderassen, cfeng stiller\}@ifi.uzh.ch, \{fabian.kuenzler\}@uzh.ch
}
}

\maketitle

\begin{abstract} 
Digitization increases business opportunities and the risk of companies being victims of devastating cyberattacks. Therefore, managing risk exposure and cybersecurity strategies is essential for digitized companies that want to survive in competitive markets. However, understanding company-specific risks and quantifying their associated costs is not trivial. Current approaches fail to provide individualized and quantitative monetary estimations of cybersecurity impacts. Due to limited resources and technical expertise, SMEs and even large companies are affected and struggle to quantify their cyberattack exposure. Therefore, novel approaches must be placed to support the understanding of the financial loss due to cyberattacks. This article introduces the Real Cyber Value at Risk (RCVaR), an economical approach for estimating cybersecurity costs using real-world information from public cybersecurity reports. RCVaR identifies the most significant cyber risk factors from various sources and combines their quantitative results to estimate specific cyberattacks costs for companies. Furthermore, RCVaR extends current methods to achieve cost and risk estimations based on historical real-world data instead of only probability-based simulations. The evaluation of the approach on unseen data shows the accuracy and efficiency of the RCVaR in predicting and managing cyber risks. Thus, it shows that the RCVaR is a valuable addition to cybersecurity planning and risk management processes.
\end{abstract}

\section{Introduction}
\label{sec:intro}
As businesses become progressively more dependent on digital technologies, their risk exposure in the cyber domain also increases. Therefore, cybersecurity must be a key concern for companies \cite{Report-WEF}. To achieve an adequate cybersecurity strategy, decision-makers (\eg business leaders and board members) must implement protection mechanisms appropriate for their risk appetite \cite{ENISA-Survey}. For that, a set of activities must be considered during the decision-making process. For example, \1 identifying cybersecurity risks and associated costs, \2 determining the impacts of cybersecurity in the business or service, and \3 understanding the business requirements and budget available for protection. These activities allow for determining a tailored strategy that satisfies technical and economic requirements to protect a company~\cite{CyberTEA}.

However, ex-ante estimations of attacks' overall financial impact are challenging due to the complexities of obtaining accurate information regarding cyberattacks and their possible risks. Cost estimations are especially demanding for Small and Medium-Sized Enterprises (SMEs) due to the absence of in-house cybersecurity expertise. Additionally, the analysis of direct and intangible costs (\ie indirect and opportunity costs) is still a challenge for decision-makers due to a set of factors, such as the information asymmetry between companies~\cite{sen2018challenges}, the lack of adequate guidelines and methods for an accurate economic measurement~\cite{ETSI-standard}, and the level of complexities added by systems and business models adopted by companies~\cite{delloite}. Therefore, there is room for novel approaches that support the analysis of economic impacts on companies, thus, enabling decision-makers to identify the optimal balance between potential losses and security investments. Such novel approaches enable even non-IT-experts to participate in the decision-making process, thus integrating cybersecurity investments into the overall enterprise risk management strategy~\cite{Enisa_SME_2021, CoreTM}.

The potential financial loss can be estimated based on companies' quantitatively measurable characteristics, such as assets at risk (\eg systems, databases, and digital businesses) and the likelihood of a successful attack happening \cite{SecRiskAI-Paper} This process can be empowered by multidisciplinary approaches that combine knowledge of both cybersecurity and finance fields to make the decision-process of investing in protections more accurate and efficient \cite{GeP}. Applying financial risk models to explain and address cybersecurity challenges is a recent trend \cite{CVaR-Survey} even though economic approaches that depict cybersecurity risk in terms of money have been established for decades \cite{anderson}. 

Two of the most accepted economic models for cybersecurity are the Gordon-Loeb model to calculate the optimal investment in cybersecurity \cite{gordon-loeb} and the Return On Security Investment (ROSI) model to compare protections based on their cost-effectiveness. Also, there are different multidisciplinary approaches exploring economic dimensions as a key pillar of cybersecurity strategies \cite{su132413677}. These approaches include frameworks to improve the understanding of cybersecurity risks \cite{BLR2017-CAS}, Machine Learning (ML) methods for assessing cybersecurity economic risks \cite{9383723, SecRiskAI-Paper}, and recommendation algorithms for cost-effective protections \cite{MENTOR}. However, current approaches still need more real-world quantitative information to understand the potential financial consequences of cyberattacks.

Theoretical models and concepts have been applied due to the need for more information regarding cyberattacks and their associated impacts. However, the lack of information negatively impacts the accuracy of these economical risk management models. One possible approach to improve the accuracy is using real-world information collected and shared by consultants and cybersecurity companies. Cybersecurity reports regarding different countries and sectors became publicly available (\eg Accenture \cite{Accenture_cost_2019}, IBM \cite{IBM_cost_2022}, Ponemom \cite{Ponemon_2012_Cost}, and Kaspersky \cite{Kaspersky_cost_2013}) in the past years. These industry reviews can provide rich data to understand a company's associated risks and costs.

Thus, this article proposes the Real Cyber Value at Risk (RCVaR) approach. It explores cybersecurity data from public reports and combines it with economic methods to predict the costs and associated deviation (\ie risks) of cyberattacks on companies. The RCVaR relies on well-placed economic methods but considers real-world information instead of only probability estimations. Based on the expected loss and risk computed using RCVaR, companies can conduct better cyber risk management and determine the best actions to maximize cost-effectiveness. The key contributions of this article are:
\begin{itemize}
    \item A methodology for extraction and statistical processing of cybersecurity-related information from industry reports;
    \item A model to estimate the potential costs of cyberattacks ex-ante based on real-world business information and security measures;
    \item The computation of the Cyber Value at Risk (CVaR) based, for the first time, on quantitative data from security reports instead of probability estimations that numerically quantify the deviation from the expected cost value; and
    \item A Web-based tool for cost and risk analysis using RCVaR straightforwardly.
\end{itemize}

The remainder of this article is organized as follows. Section II provides the relevant background for this article and reviews related work. Section III introduces the \solution{}, Section IV contains the evaluation, followed by a discussion on the identified research challenges and opportunities as Section V. Finally, Section VI concludes the article and provides details on future work.
\section{Background and Related Work} 
This section covers essential concepts for a complete understanding of the article. First, the nuances of risk management in the context of cybersecurity within companies are discussed, followed by an overview of the most common cyberattack costs. Next, the concepts of Cyber Value at Risk and its applications are introduced.

\subsection{Risk Management and Cyberattacks Costs} 
The term \textit{risk} is generally used to indicate a possibility of loss and/or damage, represented as the impact times the likelihood to happen \cite{risk-definition}. It involves some degree of uncertainty, and the resulting outcome is challenging to predict. Depending on the context, various types of risk can be found, such as business, economic, and safety risks. Also, risk can indicate the probability of an adverse event happening, such as the risk of having a higher economic loss due to cyberattacks than the one expected. Therefore, it is essential to understand that risk is an element in any company. Managing risk includes mitigating, reducing, ignoring, transferring, or even accepting when there is a reasoning behind that, especially from an economic point of view. One of the biggest risks companies face today is being the target of cyberattacks, which results in different direct and indirect costs that can disrupt a company's entire business.

Regarding the different costs of cyberattacks, the average data breach cost to Multinational Enterprises (MNEs) in 2020 was US\$ 1.09 million, compared to US\$ 1.41 million in 2019, while SMEs had to pay on average US\$ 101,000, in contrast with US\$ 108,000 in 2019 \cite{kaspersky-economics}. Businesses can reduce the cost of data breaches in different ways. For example, a quick breach detection can lower the loss by 32\% in MNEs and by 17\% in SMEs due to savings in direct and indirect costs. Also, a proactive disclosure to customers and stakeholders that a data breach has happened can lower the financial damage by 28\% in MNEs and by 40\% in SMEs.

Another factor in reducing the costs of cyberattacks is up-to-date software and hardware. If a company does not have an effective approach for updates, it could result in a 47\% increase in the cost of cyberattacks for MNEs and a 54\% increase for Small and Medium-Sized Enterprises \cite{kaspersky-economics}. In the case of ransomware affecting a company, the average cost for remediation is US\$ 761,106 in 2021 \cite{ransomware-2021}, which includes costs to mitigate, recover, and due to business disruption after cyberattacks. Payments directly to criminals to recover the data, are also becoming more frequent, with payments ranging from a few thousand to millions of dollars depending on the company's size. The average per attacked company payment made in 2021 was US\$ 233,817. The costs are also very high from a societal and governmental level, which makes the United States government push in the fight against ransomware gangs, including, in 2021, putting a US\$ 10 million reward for information that helps to catch a ransomware gang named DarkSide \cite{news-ransom}.

Besides the costs of cyberattacks, there are also many costs related to fines imposed by governments and other entities for companies that do not follow cybersecurity regulations \cite{RUOHONEN2022101876}. For example, from July 2018 to November 2021, the General Data Protection Regulation (GDPR) resulted in 837 fines in Europe \cite{gdpr-tracker}, totaling € 1.2 billion in penalties. The most common type of violation was the insufficient legal basis for data processing (299 fines), insufficient technical and organizational measures to ensure information security (179), and non-compliance with general data processing principles (178 fines).

Table \ref{tab:BackgroundCostDistribution} gives examples of the most common costs due to cyberattacks discussed by academia and industry in the last few years. These values must be considered when defining risk and the associated impacts. \textit{Anticipation} refers to costs that occur before a cyberattack has happened, including spending on additional security measures and missed opportunities due to these actions. \textit{Consequence} related costs means the immediate economic impact of a cyberattack. In contrast, \textit{Response} costs are related to expenditures that come right after the cyberattack happens, including legal, technical, and economic impacts. Thus, as highlighted in Table \ref{tab:BackgroundCostDistribution}, direct, indirect, and opportunity costs can occur in the different phases of the cyberattack, from the prevention until after the incident.

\begin{center}
\begin{table}[h!]
\caption{Dimensions of Cyberattacks Costs}
\label{tab:BackgroundCostDistribution}
\centering
\scalebox{0.67}{
\begin{tabular}{c|c|c|c} 
\hline
\textbf{Type of Cost} & \textbf{Anticipation Cost}  & \textbf{Consequence Cost} & \textbf{Response Cost} \\
\hline
\multirow{1}{*}{\textbf{Direct}} & \makecell{Cyber Insurance \\ and protections} &  Revenue Loss & Lawyer fees \\
\hline
\multirow{1}{*}{\textbf{Indirect}} & \makecell{Incident investigation \\
Employee training} & \makecell{Monitoring \\ solutions} & \makecell{Documentation \\ and report}\\
\hline
\multirow{1}{*}{\textbf{Opportunity}} & \makecell{Investment in new \\ cybersecurity strategies} &  Reputation loss & \makecell{Business \\ disruption}\\
\hline
\end{tabular}}
\end{table}
\end{center}

\subsection{Cyber Value at Risk}
Different methods, metrics, and approaches have been proposed combining cybersecurity and economic concepts to maximize companies' cybersecurity while reducing the financial losses \cite{su132413677, gordon-loeb, Enisa_ROSI, CyberTEA}. One of the most prominent for cybersecurity cost estimation is the Cyber Value at Risk (CVaR). The CVaR is a quantile-based risk metric that considers the potential harm of cyberattacks and indirectly incorporates the effectiveness of security controls. CVaR provides its users with a monetary boundary that potential costs will not surpass, with a certain probability for a specific time-frame determined by retrospective data. Initially proposed in 2015 by the Cyber Resilience Initiative of the World Economic Forum \cite{WEF_Partnering_cyber_Resilience_2015}, the CVaR traces its origins back to the Value at Risk (VaR), a statistical technique used widely in the financial sector to quantify the risk exposure within a portfolio or for a single asset \cite{orlando2021cyber}. 

\begin{figure}[ht!]
    \centering
    \includegraphics[width=\linewidth, scale=0.6]{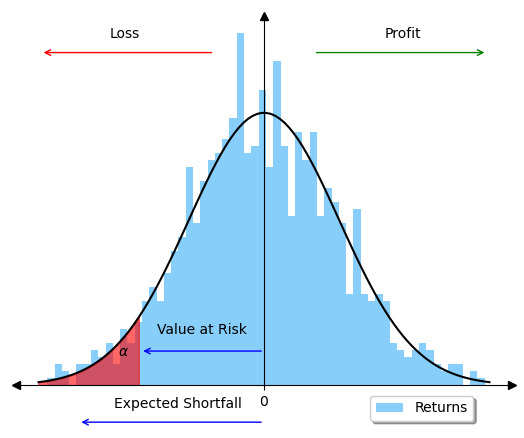}
    \caption{Value at Risk (VaR) in Stock Returns}
    \label{fig:Background_VaR}
\end{figure}

The methodology behind the financial metric VaR is key for the computation of the CVaR. According to the Efficient Market Hypothesis (EMH) \cite{fama_1970}, the returns of assets in efficient markets follow a random walk, which morphs into a Brownian motion in continuous time. Therefore, it can be argued that returns are theoretically normally distributed, as shown in Figure \ref{fig:Background_VaR}. Given this distribution, the VaR can be computed with the standard score. In this case, the VaR represents the monthly return at the $alpha$ quantile while the $1-alpha$ interval depicts the confidence.

However, cybersecurity costs cannot be assumed to have a normal distribution due to their nature and behavior, such as costs are never zero and usually follow a heavy-tailed distribution \cite{kuypers2016empirical, woods2021county}. Therefore, the CVaR has to be computed using simulations or based on historical cost time-series data. Due to its versatility, the CVaR has received particular attention from academia and industry. However, simulations needed for CVaR require computational and technical resources not often available inside companies.

Simulation-based computation requires extensive access to a company's infrastructure and data, which needs trust in the estimating authority. Additionally, simulations rely on accurate company-specific probability estimations of cyber incidents, provided by security specialists. Due to the high demand for these experts \cite{CyberTEA}, estimations can be expensive and subject to the expert's individual experience, which leads to biased probability estimates. Therefore, there is still room for approaches to computing the CVaR as a risk measure based on publicly available impact data.

Compared to traditional risk-scoring approaches in cybersecurity, CVaR offers different advantages. For example, \1 the complexity of risk can be represented by a single individually-interpretable number without assuming a uniform risk benchmark for all decision-makers. As a result, the CVaR allows decision-makers to scale risk in the cyber domain to their specific risk appetite \cite{Deloitte_CVAR_2016}. Since quantile-based risk measures are widely adopted across various industries, \2 the CVaR metric facilitates cross-domain risk comparisons (\eg financial, operational, and cyber risk). Furthermore, due to the vast number of applications of the VaR in the financial sector, multiple frameworks and models are built upon the theory of quantile risk measurements. Thus, this makes CVaR compatible with already implemented frameworks, such as those discussed in \cite{CVaR-Survey}. 

However, the CVaR model inherits not only the VaR's benefits but also its drawbacks. One limitation imposed is that the CVaR is a backward-looking measure, thus, relying only on historical data and assumptions that do not necessarily hold in the future \cite{Deloitte_CVAR_2016}. For instance, just because an event has never occurred in the past, does not mean it has a probability of zero in the future. Another often overlooked drawback is that rare events with a massive impact are not well represented by looking at past data. Therefore, the shape of the cost distribution beyond the CVaR is not captured by the quantile-based model, thus, excluding the risk of tail events.
\\ \\
Recent literature predominantly focuses on simulation-based approaches for computing the CVaR. The initial foundation was established by the World Economic Forum \cite{WEF_Partnering_cyber_Resilience_2015}, which defined essential pillars for the simulation-based methods, including \textit{Vulnerability} of assets, \textit{Assets} under threat, and \textit{Profiles of attackers} to which the assets are exposed. In a later study \cite{Deloitte_CVAR_netherlands}, Deloitte utilized these realizations to compute the expected loss and the CVaR for different sectors in the Dutch economy. Nevertheless, they do not provide a general model which applies to individual organizations. To overcome this limitation, \cite{erola2022system} proposed a refined model that calculates the CVaR based on four pillars: \textit{Asset values}, \textit{Harm probability}, \textit{Threat probability}, and \textit{Effectiveness of protection}. Once these values are determined, mainly through estimations, a Monte Carlo simulation is run to establish a distribution from which the CVaR is derived. Both studies emphasize an enormous lack of data to calculate probabilities and distributions, resulting in many probability estimations that can be skewed due to experts' behavioral biases.

Besides academia, the use of the CVaR has gained popularity in the consulting industry. With consultancy firms like MARSH Risk Consulting integrating the metric into their services \cite{MARSH_2017}, thus, providing their clients with real-world access to the model. MARSH applies a similar approach to the one outlined in \cite{erola2022system}. For that, MARSH collects relevant data at the customer's location and estimates the missing probabilities before obtaining the loss distribution through a Monte Carlo simulation. As previously emphasized, this process requires a high level of trust from the customer, as he/she must provide MARSH with critical and sensitive company data. 

Overall, the calculation of the CVaR is complex and costly (in terms of time and money). Therefore, there is still a barrier to the wide adoption of that model in the industry~\cite{orlando2021cyber}. Although the CVaR is a relatively new measure in risk management, it shows an interesting path and opportunities for multidisciplinary approaches in cybersecurity. The complexity of these approaches and the lack of data must be addressed to make adoption by companies worldwide in their risk management strategy feasible, especially for companies that do not have enough expertise, budget, and information to handle complex cybersecurity risk scenarios.
\section{\solution{} Approach}
\label{sec:solution}
\begin{figure}[ht]  
\centering
\includegraphics[width=\columnwidth]{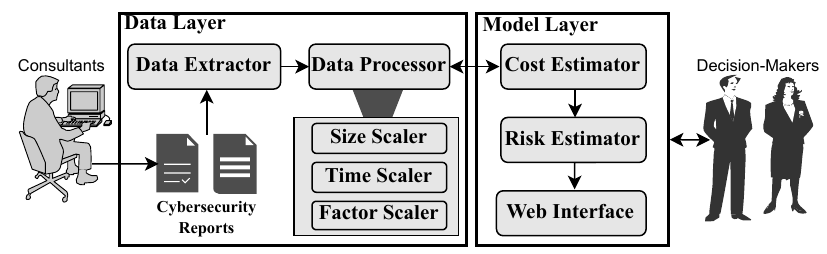}
\caption[RCVaR Approach]{Overview of the RCVaR Approach
\label{figure:RCVAR}}
\end{figure}

The RCVaR approach is proposed to assess the costs and risks a company faces due to cyberattacks. The RCVaR is the first approach of its kind to compute the CVaR with real-world data instead of estimations. For that, the RCVaR relies on concepts introduced in economic literature \cite{yamai2002validity, Deloitte_CVAR_2016} and on data from publicly available cybersecurity reports. As a quantile-based metric, the RCVaR determines the economic loss of a company with a certain degree of confidence. Thus, the costs and associated risks (\ie probability of achieving higher costs than the expected value) are predicted using a model that reflects current quantitative cyber loss behavior related to industry reports. The approach incorporates the entire process, from data extraction to cost estimation. Additionally, a Web-based platform is implemented to allow decision-makers, even those without technical expertise in cybersecurity, to apply the RCVaR intuitively and straightforwardly. Figure \ref{figure:RCVAR} highlights the conceptual architecture and workflow of the RCVaR, including its main components and relationships.

First, in the \textit{Data Layer}, inputs from cybersecurity reports are received (\eg costs per industry sectors or region). These reports can come from sources with different levels of information, such as consultancy agencies, publicly available reports, or partners. Then, the data type (\eg raw numbers, plots, or tables) is identified, and then data is extracted and processed for being used as part of the dataset. Next, the \textit{Data Processor} checks data extracted to understand all values, scales, and their relationships. Three different scaling methodologies are implemented, relying on statistical processing methods. These scalers are key for cost estimation since they put all data in the same units regarding size, time, and magnitude. The \textit{Size Scaler} individualizes the costs based on companies' valuation. A company's value is defined by its monetary equity value, which is based on the market capitalization for publicly listed companies. The \textit{Time Scaler} accounts for the time value of money. For example, A monetary amount at $t_0$ is worth more than the same nominal amount at $t_1$, given $t_0$ < $t_1$. The \textit{Time Scaler} then discounts valuation by approximated inflation numbers, while the costs are scaled with a newly developed discount factor over time. The \textit{Factor Scaler} tailors cost to a specific company based on selected business characteristics (\ie factors).  These selected business factors can also be seen as risk factors since they are used to explain the cybersecurity costs associated with companies. 

By using the features provided by the \textit{Data Processor} and its scalers, the RCVaR can then individualize the costs. For that, the \textit{Cost Estimator} computes the economic impact of a cyberattack for a company by applying Equation \ref{eq:7}, which is proposed as a core for the RCVaR approach. Furthermore, the \textit{Risk Estimator} uses the distribution of costs provided by the \textit{Data Processor} to determine the probability of incurring costs. The first step in the cost estimation process is determining the company's valuation for which the costs are estimated. The valuation should resemble the equity value of a company for the current year. Next, the valuation is discounted with the selected discount factors, where $t$ represents the number of years that have passed since the year represented by the reports (\ie information available from the reports used as input). This operation's outcome is a company's valuation for the respective year. The $cv\_ratio$ then converts this interim result to costs. Next, the costs are scaled to the year for which the estimation is required. For instance, if an approximation for 2025 is required, the $discount\_cost$ is multiplied by itself eight times (considering a report from 2017 as input). Finally, the estimation is customized by computing the product of the $param\_ratio_i$, where $i$ can be any of the 11 factors for which input can be considered (\cf Table \ref{tab:factorsNumberical}). If no specification for a factor was entered upon the execution of the estimator, the ratio is set to zero, which results in a multiplication of value one. 

\begin{strip}
\begin{equation}
\label{eq:7}
\centering
  company\_cost_{year} = \frac{{valuation_{\,ReportYear + t}}}{ discount_{\,valuation}^{t}} \times {cv\_ratio} \times discount_{\,cost}^{t-ReportYear} \times \prod_{i=1}^{11}(1 + param\_ratio_{i})
\end{equation}
\end{strip}

The two discount factors are cost and valuation. The numerical values are based on the analysis of macro-economical and cost data. By analyzing six selected reports available at \cite{RCVaR-Repository}, the cost discount factor of 9.6\% was determined as the beta of the regression conducted on the evolution of the cumulative costs over time. The same holds for the valuation discount, which was identified by running a regression on the evolution of cumulative inflation over time. The resulting beta of 1.8\% of inflation (from 2010 to 2020) can then be defined as the discount factor for the market capitalization input.

The RCVaR approach can consider different kinds of reports and types of data. Therefore, it is a generic approach that can evolve according to the needs and data availability.  An instance of RCVaR was designed and developed considering information from publicly available reports from 2010 until 2020. These reports were carefully selected according to the amount and type of data available, considering the industry sectors and the number of participants investigated. The key steps followed to define the RCVaR are described in detail in the rest of this section, including the data and statistical processing required to apply the RCVaR in a company.

\subsection{Data Sources and Cost Factors}
\label{subsec:DataSourcesAndCostFactors}
The scarcity of publicly available and rich data related to cyberattacks and breaches is a well-known challenge for cybersecurity approaches \cite{information-sharing}. This has different reasons. Firstly, companies are incentivized not to report breaches. The incentives stem from multiple factors, with the most prevalent being the anxiety of reputation loss and increased costs to raise capital. Moreover, companies usually do not want to expose themselves to liability lawsuits or signalize that they are soft targets. Secondly, assigning a monetary value to the damage caused by an attack is tough. Particularly, \textit{Indirect} and \textit{Opportunity} costs are complicated to measure. For example, it is hard to quantify the overall value lost due to customers' decision not to buy more services and goods because of the attack \cite{Cavusoglu_04}.

Consequently, there are no well-defined datasets of the monetary costs of cyberattacks publicly and widely available. However, sparse resources are accessible in reports of major consulting companies, such as Accenture~\cite{Accenture_cost_2017, Accenture_cost_2019, Accenture_cost_2021}, IBM~\cite{IBM_cost_2022}, Ponemon Institute~\cite{Ponemon_2012_Cost} and Kaspersky~\cite{Kaspersky_cost_2013}. Nevertheless, due to the different measurement units, years, and regions surveyed, it is impossible to effectively compare and merge the cost-related data between reports. Moreover, these reports only state the mean cost for their respective sample pool. The expected cost alone is not helpful due to the connection between total cost and risk (\ie the risk of deviating from the expected value). The only report analyzed that presents an array of monetary consequences of cyberattacks, from which a cost distribution, and hence risk metrics such as the CVaR, can be derived, is the Accenture report from 2017 \cite{Accenture_cost_2017}. In addition, this work investigated reports from Ponemon related to attack costs from 2010 to 2022 as possible input. Also, reports from IBM and Accenture are relevant for the RCVaR as they provide numerical cost-influential factors (\eg information regarding the influence of countries, different organization sizes or/and security systems). All cybersecurity reports analyzed are available and organized at \cite{RCVaR-Repository}.

After careful consideration, the Accenture reports~\cite{Accenture_cost_2017, Accenture_cost_2019} were defined as a primary data source for this instance of the RCVaR. Compared to other reports analyzed, Accenture is the only consulting company not exclusively focused on data breaches, meaning their reports include various cyber incident types in their surveys. Furthermore, the report covers \textit{Direct}, \textit{Indirect}, and \textit{Opportunity} economic losses of attacks. Moreover, they connect the costs to a time dimension of one year, which eases the interpretation of the cost values and enables CVaR computations. In contrast, the IBM report \cite{IBM_cost_2022} from 2022 demonstrates the cost per breach but does not provide any information about their frequency. This complicates the cost prediction since even minor impacts can have a cumulatively high cost depending on how often they occur in the investigated time frame. Thus, the Accenture reports \cite{Accenture_cost_2017, Accenture_cost_2019} are selected as the primary data sources, whereas the other consultant reports are secondary sources of information.

All the reports utilized in the context of RCVaR based their information on large interview samples with company officials. In the case of the Accenture 2017 \cite{Accenture_cost_2017} report, over 2100 managing officials of 254 companies are surveyed, which boils down to an average of 8.5 interviews per company. Most participating individuals (36\%) held key positions within their Information Technology (IT) security or operations departments. During the interviews, the participants were asked to report \textit{Direct}, \textit{Indirect}, and \textit{Opportunity} costs for each incident response over four consecutive weeks. Additionally, undisclosed shadow pricing methods were used to calculate external consequences (\eg business disruption or revenue loss). After the data was collected in the research period, the cost numbers were annualized and converted to USD according to then-current exchange rates.

Additionally, the reports serve as a valuable data source due to their diverse industry mix of companies. Most reports follow a similar sample of companies to the Accenture report \cite{Accenture_cost_2017}, which provides comprehensive coverage of 15 sectors. Most companies of the Accenture sample operate in the financial (16\%) and the industrial sector (12\%) \cite{Accenture_cost_2017}. The granularity of the sectors roughly corresponds to the 11 sectors of the S\&P 500 \cite{fidelity_sp500_sectors_22}. Regarding the company size, only large enterprises, with a minimum of 1,050 enterprise seats, were selected. The number of enterprise seats refers to the number of employees with access to internal IT systems. Regional-wise, only companies from specific western economies, such as Germany, Australia, or the United States (US), were interviewed in the 2017 report \cite{Accenture_cost_2017}.

It is crucial to outline that the study is not based on actual accounting information but on statements of multiple senior officials per company. Therefore, these statements might not accurately depict reality, even though checks were implemented in the survey to assess the correctness. Nevertheless, the overwhelming number of interviews mitigates extreme outliers' impact. Further limitations include the lack of coverage of preventive expenditures for information security and company policy measures. Moreover, the attack expenditure calculation does not consider the number of stopped attacks. It is important to note that biases (\eg non-response or sampling bias) may affect the cost estimations. Other biases can be placed, even if not identified by the report's authors. Nonetheless, the significant number of interviews and reports makes it unlikely that the cost estimates are strongly skewed. 

\begin{figure}[ht]  
\centering
\includegraphics[width=\columnwidth]{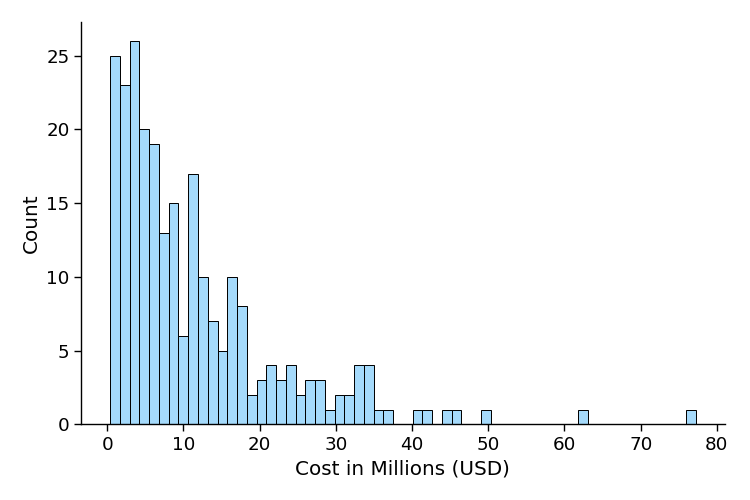}
\caption{Cost Density Distribution of the Extracted Data from \cite{Accenture_cost_2017}, based on 256 Companies Interviewed
\label{figure:original}}
\end{figure}

Extracting the single costs' data points, using the RCVaR approach, from the primary data source \cite{Accenture_cost_2017} reveals the histogram in Figure \ref{figure:original}, where the y-axis represents the number of occurrences in the sample, and the x-axis indicates the cost value associated with the observations. This tail-heavy distribution is the foundation for estimating the possible deviation from the expected value and expressing it through the CVaR. The extraction phases of the data for RCVaR rely on multiple methods from computer vision, which explored color and shape chart characteristics and human input to identify the anonymized cost numbers from plots optimally. In contrast, raw numbers were automatically extracted using Python-based scripts. All approaches work similarly by estimating the y-coordinates of the dots as well as the top and bottom lines. Once all these coordinates are known, the actual cost value can be calculated. For all computer vision-based extractions, the OpenCV library 4.6.0.66 was used. 

Furthermore, Table \ref{tab:factorsNumberical} summarizes all factors supported by the \textit{Cost Estimator}, along with their corresponding source report and the year for which cost data for each factor is available. Overall, 11 factors for customization were selected from a list of different analyzed reports. These factors are used within the RCVaR approach for explaining the cybersecurity costs exposure of individual companies, thus, facilitating cost and risk analysis based on their business characteristics.

\begin{table}[ht]
\centering
\caption{Overview of Data Sources of Factors}
\label{tab:factorsNumberical}
\begin{tabular}{l|c|c} 
\textbf{Cost Factor}  & \textbf{Years} & \textbf{Source} \\
\hline
\makecell[l]{Country} & \makecell{2017, 2018, \\ 2021, 2022}&  \cite{Accenture_cost_2017, Accenture_cost_2019, IBM_cost_2022}\\
\hline
\makecell[l]{Industry} & \makecell{2017, 2018, \\ 2021, 2022} & \cite{Accenture_cost_2017, Accenture_cost_2019, IBM_cost_2022}\\
\hline
\makecell[l]{Supplier} & 2022 & \cite{IBM_cost_2022} \\
\hline
\makecell[l]{Number of Employees} & 2017 &  \cite{Accenture_cost_2017}\\
\hline
\makecell[l]{Cloud Model} & 2021 & \cite{IBM_cost_2022} \\
\hline
\makecell[l]{Employee Training} & 2022 & \cite{IBM_cost_2022} \\
\hline
\makecell[l]{Percentage of \\Remote Employees} & 2021, 2022 & \cite{IBM_cost_2022} \\
\hline
\makecell[l]{Cyber Insurance} & 2022 & \cite{IBM_cost_2022} \\
\hline
\makecell[l]{Multi-factor \\ Authentication} & 2022 & \cite{IBM_cost_2022}\\
\hline
\makecell[l]{Identity Access \\ Management} & 2022 & \cite{IBM_cost_2022}  \\
\hline
\makecell[l]{Security Measures} &  \makecell{2017, 2018,\\ 2022} & \cite{Accenture_cost_2017, Accenture_cost_2019, IBM_cost_2022} \\
\hline
\end{tabular}
\end{table}

\subsection{Size and Time Scaler}
The estimation of the average cost per company is a challenging task, primarily due to the need to adjust the cost according to the size of the company and to determine an appropriate metric that defines the size of a company. In the domain of cybersecurity costs analysis, several approaches exist to assess the size of a company. For instance, \cite{li_2019} have utilized the number of records as a benchmark to scale the financial impacts of cyberattacks, while \cite{woods2021county} have proposed to use the company's revenue as a metric. \cite{Kaspersky_cost_2013} has categorized a company's size by the number of IT workstations, and~\cite{cavusoglu_2004} have used a company's market value to highlight the relative costs. In order to effectively match one of these metrics with the anonymized observations from the cost reports, the metric must have a publicly available, centralized and comprehensive cross-sectional track record of companies that allows sorting, aggregating and filtering companies. 

Among the metrics explored in the literature, the revenue and market capitalization meet the requirement of having enough data available, usually provided by public stock exchanges. Compared to numbers on revenue, market capitalization has the advantage of always being positive and less volatile. Therefore, market capitalization is used to represent the size characteristic of the company sample in the cost reports. However, the only actual data available on the sampled companies is the span of enterprise seats (\ie number of people with access to IT systems within a company) and the companies' sectors. As most companies are located in western economies, the average market capitalization of medium to large companies based in the US and their respective industry sectors has to be considered to approximate the market value of the analyzed sample. The US is selected because most of the companies analyzed in the report are US-based or offer services directly to the US. For that, companies in the subset of the global stock index Russel 1000 were used. This subset, called Mid-Cap Index, consists of a large spectrum of company sizes (excluding ultra-large companies) from different sectors. The conversion ratio can be computed based on the mean market capitalization of companies in the index in a given year and the average per-company cost for the same year obtained from the analyzed reports. Equation \ref{eq:6} shows how to compute this ratio, which is an essential part of the overall RCVaR formula (\cf Equation \ref{eq:7}). To further improve the robustness of the model, it is possible to compute the $cv\_ratio$ for multiple years and take the average.

\begin{equation}\label{eq:6}
  cv\_ratio_{\,\,year} = \frac{total\_avg\_cost_{\,\,year}}{avg\_market\_cap_{\,\, year}}
\end{equation}

In order to scale the valuation of the company being valued to the year of the $cv\_ratio$ and then capitalize or discount the cyberattack cost to the requested year, it is necessary to have appropriate discount factors. In consequence, RCVaR uses inflation for deprecation and capitalization of companies' valuations, while the discount factor for the size scaling is determined by regression. For that, a time-series regression of the cumulative cyberattack cost trend of the past ten years is conducted. The discount factor for cyberattack costs can be inferred from this regression's beta. Both discount factors work similarly, as shown in Equation \ref{eq:11}. The discount factors are multiplied $t$ times with the present value to scale it to the desired year. The regression on inflation rates yields a discount factor of 1.0181, corresponding to 1.8\% of annual inflation. The cost discount factor equals 1.096, which means that cybersecurity costs, on average, increase 9.6\% percent points per year.

\begin{equation}\label{eq:11}
  future\_value_{year + t} = present\_value_{\,year} \times {discount^{\,t}}
\end{equation}

Additional discount factors can be used to increase the predictability of the RCVaR model, such as the Weighted Average Cost of Capital (WACC). However, the inflation rate is considered for the RCVaR as a simplistic and well-known discount factor for real-world scenarios.

\subsection{Factor Scaler}
The factors must be computed from the report's source data after identifying the 11 significant factors for cost customization (\cf Table \ref{tab:factorsNumberical}). However, not all the factors are consistently available across all reports. For instance, while the IBM report \cite{IBM_cost_2022} numerically outlines the impact of remote workers on cost, information on this factor is absent in the Accenture reports \cite{Accenture_cost_2017, Accenture_cost_2019}. On the other hand, the number of employees is only available in the Accenture 2017 report \cite{Accenture_cost_2017}, but not in IBM's cost data \cite{IBM_cost_2022}. Further challenges arise because parameters within a factor might not align across reports. This issue is best demonstrated on the country factor, where the parameters Canada and Italy are present in the latest Accenture report \cite{Accenture_cost_2019} but not in the previous report \cite{Accenture_cost_2017}, even though the country factor as a cost driver has been researched in all of them. Thus, data cleaning is conducted to match the different reports.

Once the data is cleaned, there are still different time horizons and measurement units. This issue has to be addressed before using the information. The relative cost of a parameter (\ie business characteristic) compared to the report's overall average is computed. Equation \ref{eq:5} shows how the units can be excluded by applying a division and computing the ratio of the parameters. The parameter's relative price increase (\ie $parameter\_ratio$) as the average over $n$ reports is computed. 

\begin{equation}
\label{eq:5}
parameter\_ratio = \frac{1}{n}\sum_{i=1}^{n} \frac{cost\_param_{i} - avg\_factor_{i}}{avg\_report_{i}}
\end{equation}

This calculation can be demonstrated with a concrete example for the \textit{Banking} parameter. In 2017, the expected incident cost for companies in the banking sector was measured to be \$ 18.28  million ($cost\_param$) \cite{Accenture_cost_2017}. The average across all \textit{k} industries (parameters) is calculated with Equation \ref{eq:25}, thus, resulting in a sample average of \$ 10.348 million ($avg\_factor$), assuming companies were equally distributed across the different industry buckets. The denominator of Equation \ref{eq:5} was selected as the sample average. Its value deviates from the $avg\_factor$ due to a non-uniform distribution across categories. For the year 2017, the overall expected cost for the sample is \$ 11.7 million ($avg\_report$). 

\begin{figure}[h]
        \subfloat[Parameter Ratios of Different Regions]{\includegraphics[width=\linewidth]{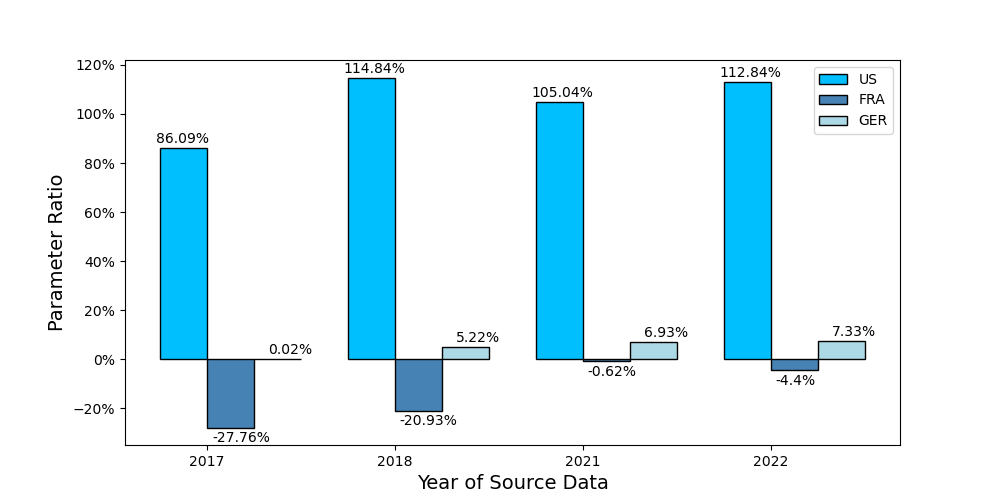}}\\
        \subfloat[Parameter Ratios of Different Sectors]{\includegraphics[width=\linewidth]{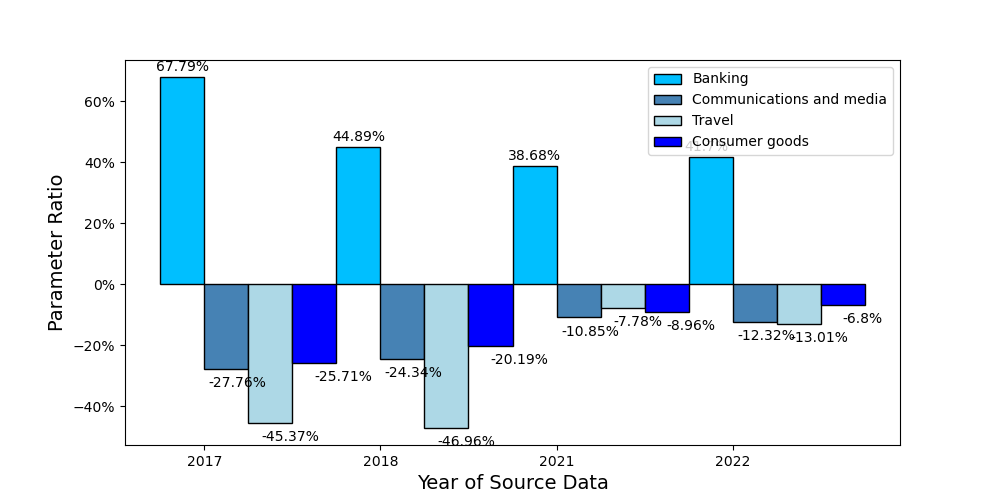}}
        \caption{Selected Parameter Ratios Within \textit{Country} and \textit{Industry} Factors}
        \label{fig:ParametersComparison}
\end{figure}

Then, substituting the numerical values into Equation \ref{eq:5}, the cybersecurity costs deviation of the banking sector compared to an average company for 2017 can be calculated. This results in a ratio of 67.79\%, which can also be highlighted in Figure \ref{fig:ParametersComparison} (b). Therefore, on average, a company in the banking sector faces costs roughly 68\% higher costs than the overall average. The analysis of the banking sector over multiple reports \cite{Accenture_cost_2017, Accenture_cost_2019, IBM_cost_2022} shows that additional costs were incurred in the following amounts: 68\% in 2017, 45\% in 2018, 39\% in 2021, and 42\% in 2022. Finally, in order to obtain a single ratio per parameter, the average of all these numbers across the \textit{n} reports in which data for that parameter is available is calculated. In the case of the banking sector, the rounded average would be 48\%. This result is then used as input in Equation \ref{eq:7} to tailor costs to companies operating in the banking sector.

 \begin{equation}\label{eq:25}
  \centering
  avg\_factor = \frac{1}{k}\sum_{i=1}^{k} cost\_params_{i}
\end{equation}

As stated in the banking sector example, the report provides two inputs for Equation \ref{eq:5}. Except for the $avg\_factor$, all variables necessary for Equation \ref{eq:5} can be extracted from the consultant reports. The $avg\_factor$ variable represents the expected value of costs if the industry is unknown. It deviates slightly from the mean sample cost stated in the report since not all industries are represented equally in the dataset. The average cost over the parameters ($avg\_factor$) is used because the deviation from the factor's average, rather than the deviation from an average company, more accurately reflects the additional costs associated with a company's sector affiliation. To compute the mean cost per factor, one calculates the average over all \textit{k} parameters within the respective factor. This computation is done using Equation \ref{eq:25}.

Prior to computing the mean of $parameter_ratios$ across reports, an examination of the ratios is conducted. Figure \ref{fig:ParametersComparison} (a) and (b) provide examples of selected parameters for the \textit{Country} and \textit{Industry} factors stated in different years (\ie different consultant reports). The ratios generally exhibit minor fluctuations over the years (\ie relatively stable). This situation can be emphasized by Figure \ref{fig:ParametersComparison} (a), where companies in the US generally experience financial costs above 86\%. In contrast, France-based companies encounter relatively lower costs than a globally average company during the analyzed time. The ranking among these three countries further stays the same for the duration of the data. 

Figure \ref{fig:ParametersComparison} (b) shows a comparable trend for the \textit{Industry} factor, wherein the industry ranking remains unaltered over time, except for the travel industry in 2021. Remarkably, even the tiny distance between \textit{Communications and Media} and \textit{Consumer Goods} is persistent over time. The relative stability of $parameter_ratios$ is particularly noteworthy, considering that the parameters were computed based on source data from different reports \cite{Accenture_cost_2017, Accenture_cost_2019, IBM_cost_2022} with different samples, measurement units, and regions. This supports the hypothesis that the relative additional (\ie company individual) costs are persistent. Thus, the average of parameters ratio (\eg sectors, regions, and security measures) over different years is an adequate approximation. 

\begin{figure}[h!]
    \centering
    \includegraphics[width=\linewidth]{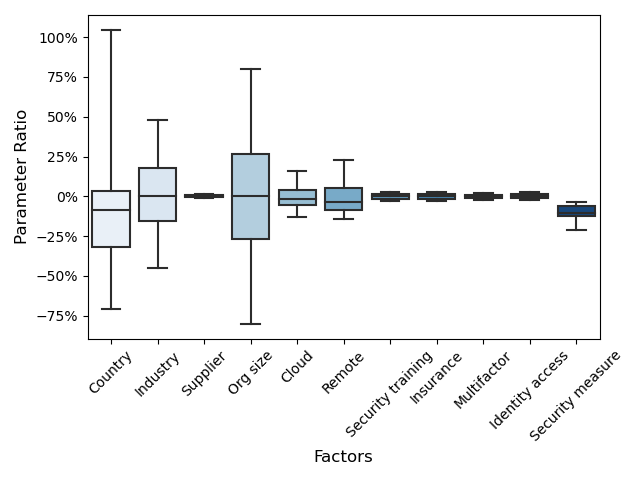}
    \caption{Distribution of the Parameters Within Factors}
    \label{fig:DistributionParameter}
\end{figure}

Further insights can be obtained by examining Figure \ref{fig:ParametersComparison} (a), which reveals that the \textit{Country} factor is characterized by a few countries (\ie parameters) with substantially higher costs than the global average. At the same time, most business locations experience relatively lower financial consequences. This indicates that some countries suffer more from cyberattacks than others. These impacts could be due to a higher frequency of attacks, more costly attack mitigation processes, or higher exposure to sophisticated attacks (\ie targeted state-side attacks). Furthermore, Figure \ref{fig:DistributionParameter} shows a symmetrical distribution of the individual additional cost (\ie $parameter_ratios$) within the \textit{Industry} factor. Comparable conclusions may be inferred by analyzing the \textit{Organization Size} (\ie Org size) factor.

A limiting circumstance of the results above is that the $parameter\_ratios$ only states the relationship of a single company's characteristic to the overall average cost. Due to data anonymity in the reports, the combined influence of multiple factors on costs cannot be explored. Therefore, uncertainty regarding the predictions persists, which can be attributed to three primary factors:

\begin{itemize}
    \item Two business characteristics might frequently appear jointly. While only one substantially influences costs, the report's analysis recognizes both as significant cost drivers. For instance, banking companies might be predominantly located in the US. In this scenario, both parameters (\ie \textit{Banking} and \textit{US}) are recognized as cost-increasing elements, even though the location might be negligible;
    \item Business characteristics can have different effects depending on whether they appear individually or in combination. For instance, the parameters \textit{Large} and \textit{Industrial} increase costs when appearing individually. However, they might have different impacts when combined, such as a higher impact on costs due to their dependencies. Therefore, if the parameters are dependent, their mutual influence cannot be expressed only by multiplying factors; and
    \item The ratios computed in this article are based on a limited dataset that spans over 5 years. More accurate representations of $parameter\_ratios$ might be obtained by more data.
\end{itemize}


\subsection{Distribution of Costs}
\label{subsection:distribution}
The data extraction process results in a data series of costs. These costs stem from different large companies operating in 15 different sectors. As depicted in Figure \ref{figure:original}, the distribution of extracted costs is heavily skewed to the left, exhibiting a long tail on the right. The heavy tail is caused by a few companies with very high annualized costs. These observations are consistent with the IBM report \cite{IBM_cost_2022} from 2020, which suggests that ``mega breaches'' (\ie incidents with exceptionally high costs) are relatively rare events. Of the 550 companies in the IBM sample, only 2.3\% experienced such significant events, roughly corresponding to 2.7\% of companies that experience extreme values (above \$ 40 million) in the Accenture sample~\cite{Accenture_cost_2017}. From the obtained data, asserting any universally valid relationship or characteristics is impossible. Nonetheless, the similarity of these findings suggests that, despite differences in years, methodologies, and companies, the distribution of extracted costs in the sample is representative of the actual distribution of costs.

These empirical results can further be supported by theoretical reasoning: Cyber criminals do not typically target specific victims \cite{shinder2008scene}; instead, they are drawn to victims who respond to bait or have vulnerabilities. To explore the implications of this hypothesis, we examine Phishing attacks, which are reported to be the second most common and the most costly attack vector according to the IBM cost report of 2022 \cite{IBM_cost_2022}. However, only a fraction of the Phishing attempts will result in a successful attack and, therefore, in financial gain for the attacker. If one plotted the emails sent to possible victims on the y-axis and the monetary gain of the criminal per email on the x-axis, the heavy-tailed distribution would closely resemble the one shown in Figure \ref{figure:original}. As the profits of cyber criminals are directly related to the victims' costs, the distribution must be mirrored to get the company's cybersecurity costs. This can be accomplished by relabeling the x-axis from gains to cost. If all other attack vectors are subject to the same distribution, the overall distribution is a linear transformation of the previously established Phishing cost distribution.

Upon extracting the data, a goodness-of-fit test is conducted to determine which distribution fits best the sample from the report. The test is necessary to deduce the continuous function from which the $\alpha$ quantile can be computed. The Kolmogorov-Smirnov (KS) 1-sample test is used to perform the goodness-of-fit test. Initially, the data is arranged in ascending order. Subsequently, the Empirical Cumulative Distribution Function (ECDF) is generated and compared against the target distribution's Cumulative Distribution Function (CDF). Lastly, the vertical distance of each data point to the reference CDF is calculated. The best-fitting distribution is the one with the lowest single maximum distance \cite{massey1951kolmogorov}. 

\begin{figure}[ht]
    \centering
    \includegraphics[width=1\linewidth]{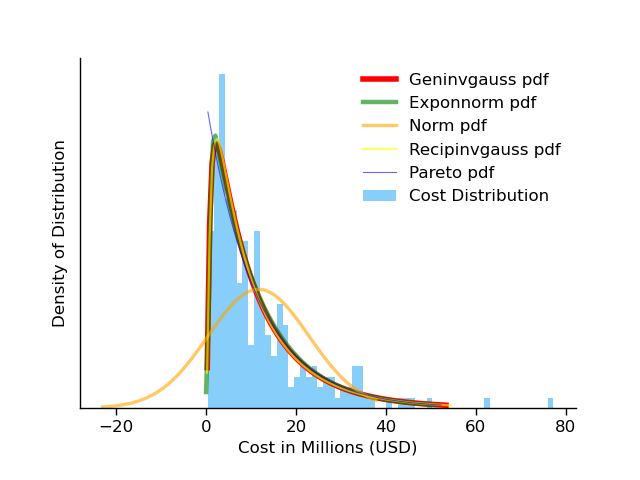}
    \caption{Distributions Fit to Sample}
    \label{fig:KS1}
\end{figure}

The performed tests suggest that the Generalized Inversion Gaussian (Geninvgauss) distribution is least likely to be rejected. Important to point out is that the two statements: ``distribution which is most likely not to be rejected'' and ``most probable distribution'' are not equivalent. Therefore, it is impossible to identify ``the one-size-fits-all'' solution. This circumstance can also be observed in Figure \ref{fig:KS1}, where only minor differences between the distribution with the highest p-value and the subsequent ones are visible. Additionally, some assumptions of the KS-test had to be relaxed in order to conduct the evaluation. An elaborate discussion on the specific use-case can be found at \cite{Kunz23}. Table \ref{tab-p} shows that the Geninvgauss is the assumed best fit for RCVaR due to its p-Value.

\begin{table}[ht]
\caption{p-Values of Analyzed Distributions}
\label{tab-p}
\centering
\begin{tabular}{l|l}
\hline
\textbf{Distribution} & \textbf{p-Value} \\ \hline
Geninvgauss           & 0.98921          \\ \hline
Exponnorm             & 0.96195          \\ \hline
Norm                  & 4.89e-6          \\ \hline
Recipinvgauss         & 0.97147          \\ \hline
Pareto                & 0.46933          \\ \hline
\end{tabular}
\end{table}

The distribution of costs is derived from the loss of multiple companies. Consequently, the data does not represent a time series of a single company. Thus, the Geninvgauss distribution is inferred by a data snapshot of multiple companies with different characteristics, which poses an issue when applying the distribution to a single company to assess its risk outlook. Since there are proven similarities \cite{akkiraju2010discovering} among larger corporations, it can be reasonably assumed that the cost distributions between a time series of a single company and a snapshot of numerous companies vary insignificantly. Thus, the established distribution can be used cautiously to infer a company's CVaR.

\subsection{Web Platform}
The source code of RCVAR and a dataset of hypothetical companies and their costs are accessible at \cite{RCVaR-Repository}. A running version of the platform is also available at \url{https://www.csg.uzh.ch/rcvar/}. The front end was developed using Meta's React Framework in conjunction with Typescript, whereas the back end employs Flask with Python for the RESTful API. The platform was designed and implemented with extensibility in mind, thus enabling the swift integration of new models and factors.

\begin{figure}[ht]
    \centering
    \includegraphics[width=1\linewidth]{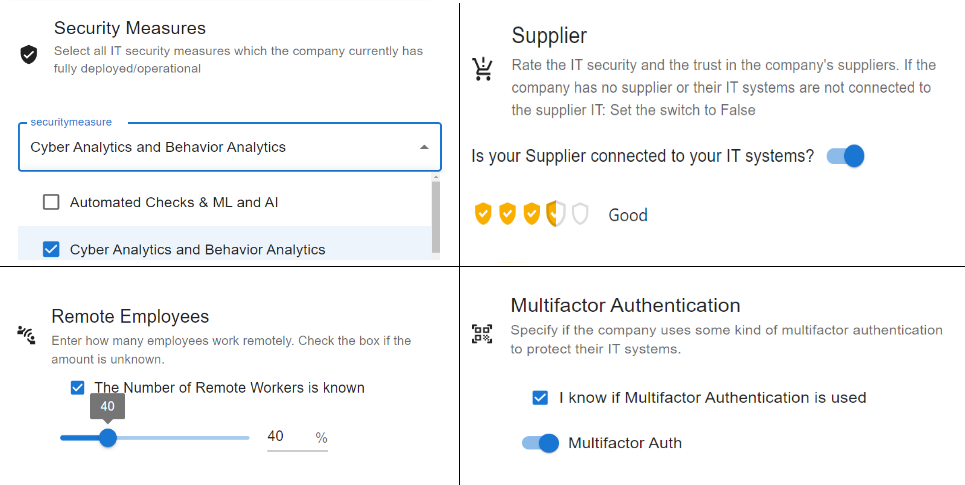}
    \caption{Web-based Interface for Input Factor Configurations}
    \label{fig:SummaryInputs}
\end{figure}

\begin{figure}[ht]
    \centering
    \includegraphics[width=1\linewidth]{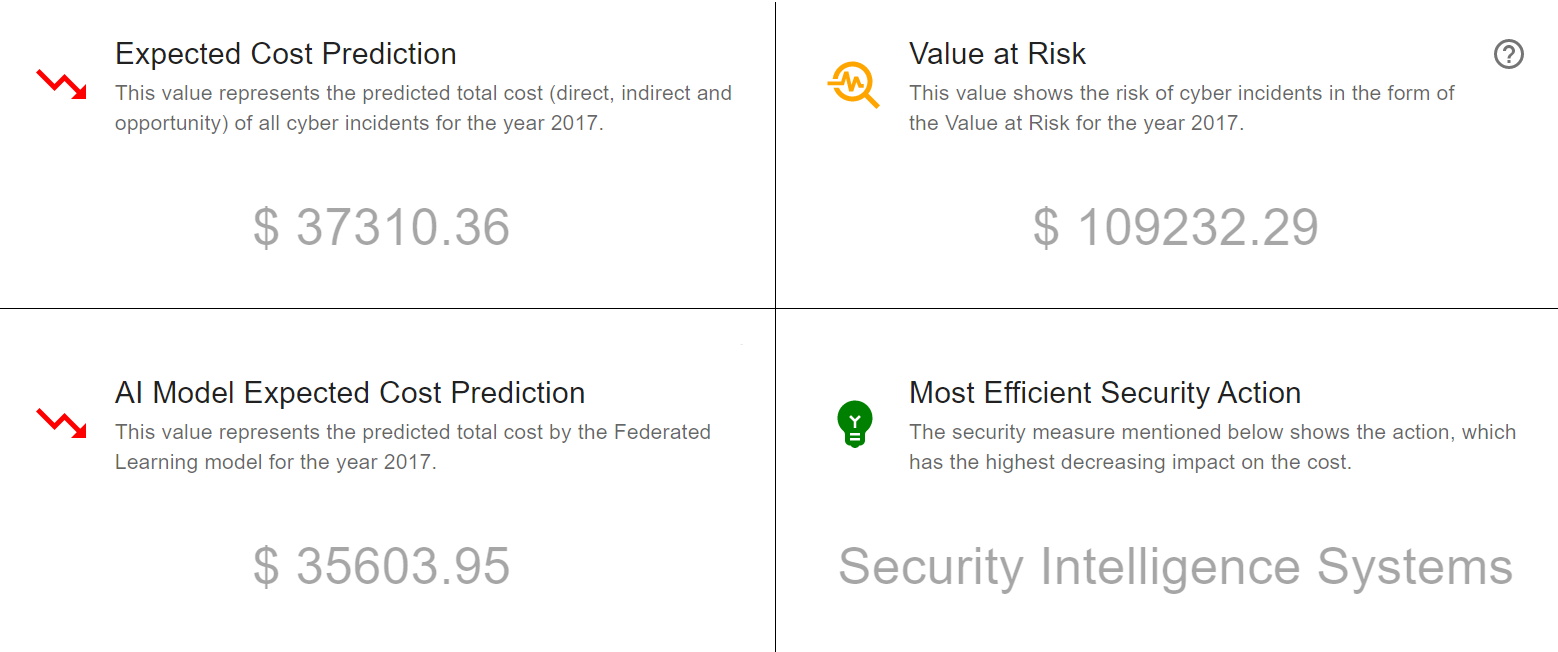}
    \caption{Dashboard of RCVaR Outputs}
    \label{fig:SummaryCosts}
\end{figure}

Figure \ref{fig:SummaryInputs} highlights examples of four out of eleven (\cf Table \ref{tab:factorsNumberical}) factors (\eg \textit{Security measures}, \textit{Suppliers}, \textit{Remote employees}, and \textit{Authentication measures}) which are taken into account for configuring a company's cost prediction. On the other hand, Figure \ref{fig:SummaryCosts} illustrates the resulting costs generated by the applied equations and how they are presented to the end-users. The expected cost prediction of the RCVaR is displayed in the top left corner, followed by the CVaR metric for the company on the right. Also, a supervised ML-based model prediction is shown in the bottom left. This model was trained in a federated learning approach on a synthetically generated dataset that simulates the cost patterns observed in the reports \cite{Kunz23}. Furthermore, the platform can suggest the most effective security action for reducing expected costs. This feature can potentially be enhanced by incorporating the recommender engine proposed in \cite{MENTOR} in future iterations.

It is noteworthy that the RCVaR assumes constant variance of the cost distribution over time and factors. Consequently, companies with varying business characteristics are assumed to have the same variance and, therefore, the same risk of deviating from the expected value. This means the individualized CVaR is always 2.9 times the expected value, with a 95\% confidence level. This circumstance can also be observed in Figure \ref{fig:SummaryCosts}. Given that more data becomes available in the coming years, the variance change due to business characteristics can be investigated, ultimately leading to more accurate predictions of the RCVaR.

The RCVaR approach can be summarized as follows: A company's valuation is initially established using current economic indicators. Subsequently, the distinctive attributes of the company are identified and inputted together with the valuation into Equation \ref{eq:7} to predict the expected cost. Next, the risk in the form of the CVaR is deduced from the continuous Generalized Inverse Gaussian distribution. In order to generate the probability density function (pdf) of the distribution, the initial parameters from the sample are inputted in the respective function of the SciPy library, except for \textit{scale} and \textit{loc}. The scaling of the distribution necessitates modification of these two parameters. To determine the new location, the original variable is scaled proportionally to the size of the expected value. Thus, the distribution is shifted on the x-axis until the correct position is attained. Subsequently, the scale is adapted similarly. These steps culminate in a linearly scaled version of the original distribution. Additional details and discussions regarding RCVaR can be found at \cite{Kunz23}. The final results of these computations (\ie RCVaR model) are the predicted expected cost value as well as the risks for a specific company.

\begin{figure}[htpb]
    \centering
    \includegraphics[width=0.52\textwidth]{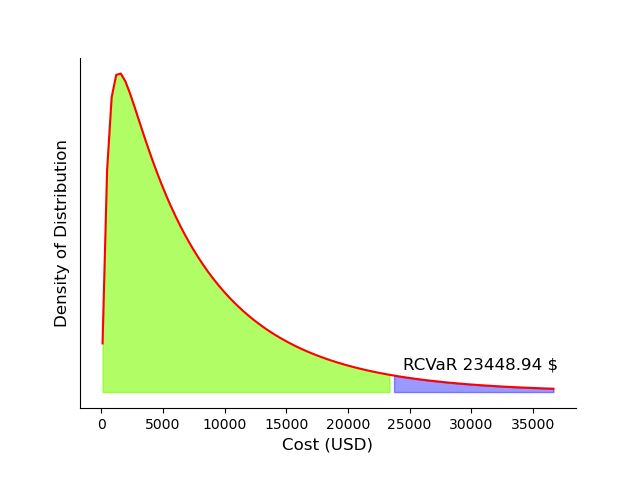} 
    \caption{Example of RCVaR of a Company in 2019 with 95\% Confidence}
    \label{fig:ValueAtRiskExample}
\end{figure}

Figure \ref{fig:ValueAtRiskExample} shows an example of the scaled density function. It shows the distribution of costs of a company with a valuation of \$ 1,000,000 for the year 2019. Based on this distribution, the discrete value to which the realized costs are less or equal, with a certain probability (95\%), is calculated. This discrete value can be found at the 95\textsuperscript{th} quantile in this example. The green area under the curve marks the range of costs up to the 95\textsuperscript{th} quantile, while the x-axis of the blue area shows costs that occur with a probability of 5\% or less. Thus, the expected costs are determined to be rough \$ 8,000 with a CVaR being equal to \$ 23,448. 
\newpage
\section{Evaluations}
\label{sec:evaluation}
The evaluation focuses on the cost prediction and risk components of the RCVaR. First, the monetary impact projected by the RCVaR is contrasted with the cost predictions from out-of-sample academic and industrial research. Next, the RCVaR's ability to capture risk (\ie likelihood of costs beyond the expected cost) is discussed by comparing the cost distribution adopted in the model with others provided in the literature.

\subsection{Cost Estimation}
As part of the assessment, the cost predictions generated by RCVaR are contrasted with the actual cost estimates reported in the relevant literature, such as publicly available surveys and reports. However, due to the anonymized nature of the data, it is challenging to identify the business attributes associated with the cost estimation in the relevant literature. Therefore, the quantitative evaluation displayed in this chapter only focuses on three separate predictions \cite{Kaspersky_cost_2013, woods2021county, Deloitte_2016}. None of these reports were part of the cost estimator's development. Thus, the experiment can be viewed as an out-of-sample assessment, indicating that the RCVaR model had no access to this data before the evaluation to prevent bias.

The first cost estimation number considered from the literature stems from the Kaspersky report \cite{Kaspersky_cost_2013} published in 2013. Kaspersky carried out approximately 3000 interviews with IT specialists familiar with IT security and the business process in their companies. The complete sample comprised 2364 companies, which were categorized into two groups depending on their amount of digitized workplaces. Companies with less than 1500 computerized workplaces are labeled as SMEs, while the others are classified as large corporations. According to the report analyzed, the average loss incurred by SMEs due to cyber incidents was \$ 50'000 for the year 2013.

The mean equity value of an SME for the year 2022 can be approximated by using the Venture Capital (VC) data provided by Pitchbook \cite{Pitchbook_report_22}. Combining this information with the year of the Kaspersky study, all the fundamental data necessary to calculate the expected yearly cost is given. Upon entering the target year 2013 with a market capitalization of \$ 168 million, the model discounts the equity to the year 2017 before transforming it into the cost (\ie Size Scaler) and further discounting it to the desired year (\ie Time Scaler). As per the calculation performed by the RCVaR model, the costs amount to roughly \$ 70,000. This value can be observed in Table \ref{tab:QuantitativeEvaluationUseCase}. Additionally, the table highlights the deviation between actual and estimated value, which is equal to 44\% for the Kaspersky scenario.

This deviation may indicate an inaccurate estimation. Nevertheless, it is crucial to point out that the value of \$ 50,000 from the Kaspersky's report reflects the average cost per severe incident. In contrast, the RCVaR model's cost prediction represents annualized incident costs. Given that companies have multiple severe incidents per year on rare occasions, annualized costs of \$ 70,000 are plausible. Furthermore, the VC equity input might lead to overestimating the cost in this specific instance since VC valuations rose to an all-time high in the first quarter of 2022. Using a smoothed market valuation (\ie the average over the past 12 months), the equity approximation of SMEs equals \$ 134 million. After re-running the estimation, this new input yields annualized anticipated costs of \$ 57,426, which is very close to the original value obtained by Kaspersky. 

\begin{center}
\begin{table}[h]
\caption{Out-of-Sample Cost Estimations Evaluation by Reports}
\label{tab:QuantitativeEvaluationUseCase}
\centering
\scalebox{0.65}{
\begin{tabular}{c|c|c|c} 
\hline
\textbf{Cost Source}  & \textbf{\makecell{Cost Estimation \\ of Source}} & \textbf{\makecell{Cost Predicted \\ using RCVaR}} & \textbf{\makecell{Absolute Percentage \\ Error}} \\
\hline
\textit{Kaspersky Report \cite{Kaspersky_cost_2013}} & \$ 50,000  &  \$ 71,997 & 44\% \\
\hline
\textit{Woods et al. \cite{woods2021county}} & \$ 428,000 & \$ 391,534 & 9\%\\
\hline
\textit{\makecell{Deloitte - Health Insurer\\ Use Case \cite{Deloitte_2016}}} & \$ 465,333 & \$ 711,141 & 53\% \\
\hline
\end{tabular}}
\end{table}
\end{center}
\vspace{-1.1cm}

The subsequent study by Woods et al. \cite{woods2021county} used pricing data from roughly 7000 observations across 26 insurance companies to derive a real-world cost distribution. Based on this foundation, the authors predicted the price of a hypothetical Californian retail company. The retail company has a yearly revenue of \$ 50 million. Thus, the Return on Equity (ROE) ratio approximates the company's valuation by converting revenue to equity. The ROE ratio is a percentage measure representing the revenue ratio in terms of the money invested to achieve this income. 

The NYU \cite{ROENYU} provides a comprehensive overview of ROE ratios for different regions and industries. This ROE data can be used to convert the revenue to equity, thus, resulting in a market capitalization of \$ 249 million for a US-based general retail company for the year 2021. Adjusting for inflation, the retail company would have a value of \$ 253 million for 2023. Providing the industry (Retail), the equity value (\$ 253 million), the location (USA), and the desired year (2021). All the information needed to analyze the company with the RCVaR is given. 

The RCVaR output and the result from the study presented in \cite{woods2021county} provide similar values, which strongly indicates the model's accuracy. Both values are annualized, and more data regarding the company's features were available to produce the RCVaR's prediction. In another scenario, a higher percentage error is obtained by replicating the hypothetical scenario described in Deloitte's cybersecurity cost strategy study \cite{Deloitte_2016}. In this scenario, a health insurance company faces direct and indirect costs over five years due to a data leak. However, similar to the Kaspersky report \cite{Kaspersky_cost_2013}, Deloitte's cost number is representative of a single incident rather than an annualized loss. Comparing the results of \cite{woods2021county}, \cite{Deloitte_2016}, and \cite{Kaspersky_cost_2013}, it can be concluded that the RCVaR becomes more accurate as more information is provided to it. Overall, the RCVaR demonstrates proficiency in its intended use of forecasting annualized cost projections. However, the accuracy of predicting single incident costs is relatively low. Further evaluations conducted for the RCVaR \cite{Kunz23} show that the output has a strong real-world connection. 

Table \ref{tab:QuantitativeEvaluationUseCase} shows that the RCVaR can be used to approximate the cost numbers of three reports with different methodologies, regions, years, and information sources. Nevertheless, additional testing is necessary to prove the real-world validity of discount and cost-increasing factors within the RCVaR approach. The RCVaR yields accurate annualized predictions when tested on ``unseen'' data. Even with limited input information, the model still provides reasonable estimates within the relevant range. Initial evidence suggests that the model's ability to reflect real-world cost patterns instead of being purely theoretical. Additional evaluations of RCVaR are also available at \cite{Kunz23}. Thus, based on public information extracted from industry reports, the proposed RCVaR model achieves highly accurate cost predictions on theoretical and real-world companies.

\subsection{Risk Estimation}
In order to assess the risk estimation capacity of the RCVaR, the proposed General Inverse Gaussian (GenInvGauss) ``risk'' function is compared to cost distributions suggested in prior research. As per \cite{erola2022system}, both Log-Normal and Skew-Normal distributions exhibit a satisfactory fit for actual distributions, although other options may also prove effective. Nevertheless, it is widely considered inevitable that a heavy-tailed distribution characterizes the financial impact of cybersecurity incidents, as supported by \cite{kuypers2016empirical, woods2021county}.

\begin{figure}[h]
        \subfloat[Cost Distribution Obtained from MARSH \cite{MARSH_2017}]{\includegraphics[width=0.55\linewidth]{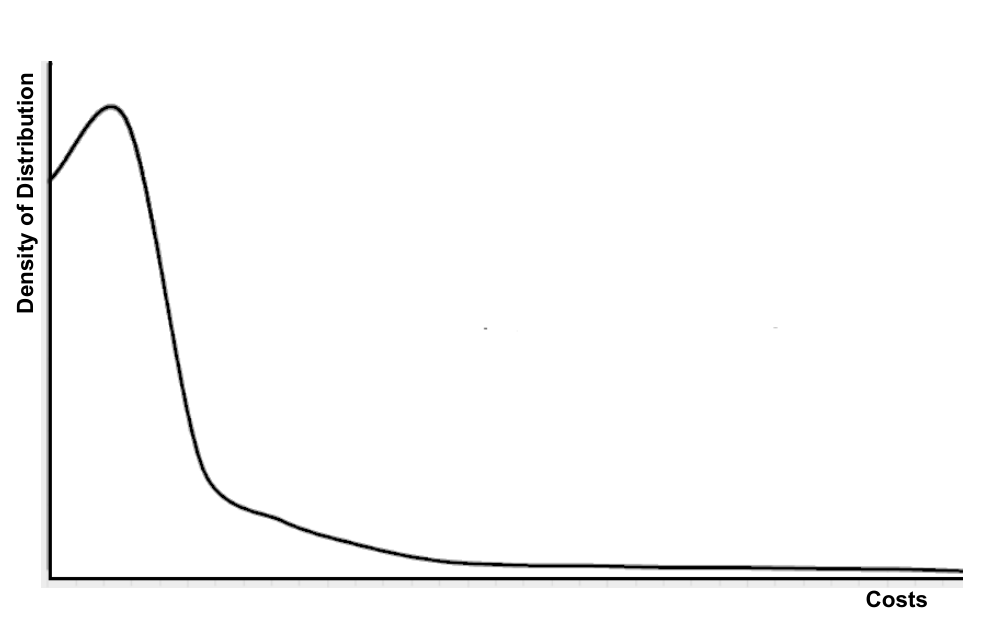}}
        \subfloat[Cost Distribution of the RCVaR]{\includegraphics[width=0.45\linewidth]{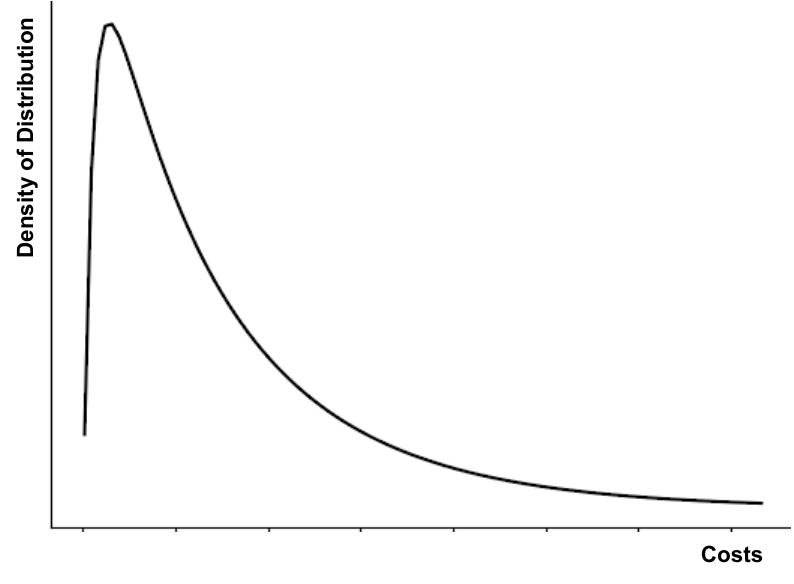}}
        \caption{Comparison of the cybersecurity cost Density Distributions of MARSH and RCVaR}
        \label{fig:RiskEvaluationGraphs}
\end{figure}

The literature includes additional distributions such as Power-Law, Pareto, and Weibull. It is relevant to emphasize that none of these distributions could be rejected through the Kolmogorov-Smirnov test conducted in the scope of this work (\cf Section \ref{subsection:distribution}). Notably, the Pareto distribution even had the fourth highest p-value after the GenInvGauss. Given the similarity between these distributions, it can be assumed that the risk measure of the RCVaR is a reasonable approximation of real-world circumstances. The key difference between distributions in the cyber economics research is whether cost density at the origin is zero. Since the RCVaR accounts for all cybersecurity incident costs (\ie direct, indirect, and opportunity costs), costs are assumed to be non-zero even if protective measures thwart cyberattacks. As a result, bell-shaped curves (\eg Weibull, Skewed-Norm, and GenInvGauss) are deemed more appropriate.

In a final comparison, the risk distribution of the RCVaR is assessed against a distribution generated by MARSH through a Monte Carlo simulation. MARSH employs this process to evaluate a company's risk (\ie CVaR) and advise its clients on cybersecurity planning. Figure \ref{fig:RiskEvaluationGraphs} (a) illustrates the distribution established by MARSH \cite{MARSH_2017}, while Figure \ref{fig:RiskEvaluationGraphs} (b) depicts the distribution of the RCVaR. Both figures have similar shapes, which suggests that the RCVaR produces reasonable CVaR results. However, the RCVaR has density zero at the origin, which can be explained by the different definitions of cyberattack costs covered. Examining the matching distributions in the literature and the cost distribution of the RCVaR, it can be concluded that the RCVaR provides a relatively close approximation of the real-world cost distribution, thus providing accurate CVaR risk results using real-world reports.

\section{Discussion}\label{sec:discussion} 
Companies and institutions (\eg WEF and FAIR Institute) are exploring and discussing the opportunities and challenges surrounding economic metrics for cybersecurity. In this direction, there are efforts to establish standardized methods for calculating the CVaR, such as the one proposed by the FAIR institute \cite{FAIR}. FAIR emerged in 2018 as an international board aiming to establish itself as the authority for the CVaR model development and its practical use by providing best practices for companies and products that want to apply CVaR in real-world scenarios. One of the biggest challenges is still real-world data. While FAIR still relies on simulations and specific solutions, RCVaR emerges as an alternative focused on real-world industry reports.

However, the RCVaR's risk and cost estimation have limitations and assumptions. The main assumption is that the data obtained from surveys and reports accurately reflect the actual loss numbers. Although the consultant agencies behind the reports (\eg Accenture and IBM) introduced checks and balances during their surveys, the possibility of selection bias cannot be entirely removed. Furthermore, even though the results of the experiments are adequate, data pollution could have occurred during the data extraction phase.

Due to data anonymization, only the relationship between one factor and costs can be observed in a timeframe. However, cross-correlation effects may arise. For instance, the \textit{Industrial} and the \textit{No Identity Access Management} parameter increase the costs when considered individually. Their mutual impact may not necessarily result in the same outcome. However, the diverse data sample and careful factor selection suggest that the RCVaR output is unlikely to be significantly impacted due to cross-correlation effects. Additionally, the anonymization presents challenges in linking a company to its valuations, with the mean capitalization of the Russell Mid-Cap index serving only as an approximation of the actual company value. Thus, the predicted costs remain sensitive to the market capitalization input used by the RCVaR model. Therefore, every cost prediction should be subject to a common sense review to ensure better results. Additionally, incorporating data from future reports can increase the robustness of the RCVaR's result.

Furthermore, the model often relies on simplified economic concepts. More data or more accurate models can address this limitation. For example, the model currently uses inflation as the valuation discount factor instead of the more sophisticated concepts. Furthermore, the applicability of the RCVaR model to non-listed SMEs remains to be determined, since data regarding cybersecurity costs for SMEs is not extensively available. Nevertheless, limited quantitative tests suggest that the RCVaR can provide rough estimations for SMEs. 

ML techniques further offer promising research opportunities because of their ability to learn from datasets, generate knowledge, and understand patterns of a given input (\eg business profile and assets). Approaches like the RCVaR can be used as input for a supervised ML model. For this purpose, the web-based interface already provides a comprehensive dataset of hypothetical companies, incorporating the cost behaviors observed in the consultant reports. Furthermore, the RCVaR implementation provides, as a Proof-of-Concept, a deployable federated learning model to address the issue of data scarcity, with particular consideration of privacy concerns. Thus, ML-based solutions (\ie Federated Learning) are potential allies for specific cybersecurity planning tasks and measurements to circumvent current limitations and even enable more collaborative approaches to increase the accuracy of the measurements. Additional evaluations and details of the RCVaR approach can be found at \cite{Kunz23}, including an analysis of the potential benefits of ML models based on RCVaR.
\section{Conclusions and Future Work}
\label{sec:conclusions}
The RCVaR approach introduces clear steps for processing data sourced from cybersecurity reports and consultant agencies. By relying on extracted information, the RCVaR is able to calculate the expected costs and economic risks associated with cyberattacks. For that, the RCVaR explored and addressed the limitations of the state-of-the-art CVaR metric and introduced a new methodology to compute the expected cost. It enhances the practical application and accessibility of cybersecurity economic models for risk and cost predictions.
 
The evaluations of RCVaR demonstrate its potential to conduct cost and risk estimations on real-world organizations. Its novelty consists of \1 Refining the overall risk estimation approach by relying on real-world analysis and surveys publicly available instead of just simulations and \2 it identifies main cost drivers and combines them in a model to accurately predict expected annualized costs due to a variety of cybersecurity incidents. Due to its practical nature supported by theoretical models, the RCVaR approach can serve as a Key Performance Indicator (KPI) and Key Risk Indicator (KRI) in real-world companies. By leveraging the RCVaR approach, businesses can effectively manage cybersecurity planning and risk management tasks. Furthermore, the RCVaR approach can be viewed as the foundation for developing novel techniques for risk assessment.

Future work includes additional evaluations of RCVaR in real-world scenarios as well as further calibration of the factors according to specific companies' demands. Especially the development of approaches to scale the variance of the cost distribution due to different factors and time might be of strong interest, such as the inclusion of implied volatility of cybersecurity stock indexes. Also, additional factors can be included in the approach if more data is available regarding cyberattacks, thus, not relying only on reports but also on data collected by Security Information and Event Management (SIEM) systems deployed in companies. Furthermore, ML-based applications of the RCVaR can be considered to refine estimations provided by the RCVaR.

\section*{Acknowledgements}
This work was supported partially by \textit{(a)} the University of Z\"urich UZH, Switzerland and \textit{(b)} the European Union's Horizon 2020 Research and Innovation Program under Grant Agreement No. 830927, the CONCORDIA project.

\bibliographystyle{IEEEtran}
\balance
\bibliography{bib/main.bib}

\end{document}